# Geometric Metasurface Fork Gratings for Vortex Beam Generation and Manipulation

*Shumei Chen, Yuan Cai, Guixin Li\*, Shuang Zhang\*, and Kok Wai Cheah\**

[1]Department of Physics, Hong Kong Baptist University, Kowloon Tong, Hong Kong
[2]Materials Characterisation & Preparation Facility, The Hong Kong University of Science and Technology, Clear Water Bay, Kowloon, Hong Kong, Hong Kong
[3]School of Physics and Astronomy, University of Birmingham, Birmingham, B15 2TT, United Kingdom

\*Corresponding Author: E-mail: (ligy@bham.ac.uk; s.zhang@bham.ac.uk; kwcheah@hkbu.edu.hk)

In recent years, optical vortex beams possessing orbital angular momentum have caught much attention due to their potential for high capacity optical communications. This capability arises from the unbounded topological charges of orbital angular momentum (OAM) that provides infinite freedoms for encoding information. The two most common approaches for generating vortex beams are through fork diffraction gratings and spiral phase plates. While realization of conventional spiral phase plate requires complicated 3D fabrication, the emerging field of metasurfaces has provided a planar and facile solution for generating vortex beams of arbitrary orbit angular momentum. Among various types of metasurfaces, geometric phase metasurface has shown great potential for robust control of light and spin controlled wave propagation. Here we realize a novel type of geometric metasurface fork grating that seamlessly combine the functionality of a metasurface phase plate for vortex beam generation, and that of a linear phase gradient metasurface for controlling the wave propagation direction. The metasurface fork grating is therefore capable of simultaneously controlling both the spin and the orbital angular momentum of light.

## 1. Introduction

Orbital angular momentum (OAM) of photon, as first proposed by Allen et al. [1], is manifested as a helical phase front: $\exp(il\varphi)$, where $l$ and $\varphi$ are the topological charge and azimuth angle respectively. It has attracted a broad range of interests from optical community [2-4] and has been extensively used in macro-manipulation [4, 5], quantum optics [6-8], optical communications [9] etc. OAM of light has unlimited number of eigen-modes (with $l$=1, 2, 3...), so it offers much more information capacity than polarization state does for optical communication applications in both the classical and quantum optics regimes [6-9]. Of particular interest is the interaction between SAM and OAM of light by using the 'q-plate' (liquid-crystal based spatial variant birefringent plate) [10]. The optical spin-orbit interaction through 'q-plate' can be used to simultaneously control OAM by changing incident states of SAM. Alternative approaches for generating coupled SAM-OAM of light are based on ring resonator [11] or the liquid crystal based spatial light modulator (SLM) [3, 12, 13]. While these methods have been well established for both visible and near-infrared wavelength applications, they inevitably require a large number of conventional optical components such as polarizers, SLMs, quarter wave-plates, beam splitters and so on. Therefore, the complex optical setup in the conventional methods limits the development of on-chip device integration.[14-16]

On the other hand, with the development of plasmonics and meta-materials [17-21], light-matter interaction can be artificially manipulated or enhanced by using metallic nanostructures with judiciously engineered electromagnetic response. For example, generation of optical vortex has been demonstrated by using plasmonic metamaterials [22, 23]. Very recently, plasmonic metasurface, a two-dimensional meta-material consisting of arrays of subwavelength plasmonic antennas for generating phase discontinuity, have been proposed for arbitrarily manipulating the wavefront of light [24]. Recently, various optical phenomena from metasurface devices, such as anomalous refraction and reflection, flat axicon lens, three dimensional hologram, optical spin hall effect, and generation of optical vortex beams (OVs) were demonstrated [25-37]. In this work, we circumnavigate the constraint of conventional ways of generating SAM-OAM of light by using ultra-compact metasurface fork gratings.

This fork-type plasmonic metasurface consists of spatially variant nanoslits on ultrathin metal film, serving as a highly compact device for both generating OVs with various OAM states, and splitting these OVs to different diffraction orders according to their SAM states. Unlike the conventional diffractive optical element [38, 39], the metasurface proposed here enables local manipulation of phase with pixel size down to the sub-wavelength scale. The metasurface fork grating in this work is designed for circularly polarized light, the appropriate design of phase gradient of the fork grating canand capable of separatinge light with different SAM-OAM modes to either 1st or -1st order. In comparison, most of the similar previously demonstrated metasurface devices [30, 35-37] only generate the optical vortex while lacking the ability to separate the coupled SAM-OAM modes of light. Thus we can take the advantage of holographic metasurface and develop a new solution to generate coupled SAM-OAM of light based on a spatial projection scheme.

## 2. Experimental Methods

Based on the principle of optical holography [40, 41], the fork-type phase distribution: $\varphi(x, y) = (2\pi x / P) + q\phi$, where $q$ is the number of phase singularities and $\phi = \arctan(y/x)$, is encoded into metasurface by using nanoslits with various orientation angle $\theta$ respect to x-axis (**Figure 1**). For circularly polarized incident laser, the transmitted light with opposite handiness has a geometric Berry phase factor: $\exp(2i\theta)$ [42]. By changing the orientation angle of the nanoslit from zero to π, a phase distribution ranging from zero to 2π is obtained. In this work, the nanoslits with variable rotation angles ($\theta$ from 0 to 7π/8) are milled by using focus-ion-beam method into a 80 nm thick aluminum thin film, which is deposited on glass substrate through thermal evaporation process. Each period of the metasurface fork grating consists of eight unit cells with total period $P$ of 3.2 μm. The interaction between laser with Gaussian beam profile and metasurface fork grating is then experimentally demonstrated. A He-Ne laser beam with horizontal (H-) polarization state is normally incident onto the metasurface fork via a lens with low numerical aperture (N. A. = 0.18).

**3. Results and Discussion**

As is shown in **Figure 2**, the optical vortex with different OAM values are projected to ± 1$^{st}$ diffraction orders. For the fork phase distribution with topological charge: $q = 2, 3$ (F2 and F3), the OAM mode of transmitted optical vortex has a topological number: $l = \pm 2, \pm 3$ respectively, which has been verified by the interference patterns of light with OAM in previous studies. Since the radius of the donut patterns of optical vortex is proportional to $\sqrt{l}$, the topological charge can be indirectly checked by calculating the radius ratio $\delta$ of two vortex beams with different $l$. For the metasurface devices F2 and F3, the measured value of $\delta$ is ~ 0.807, which agrees well with the theoretical value: $\sqrt{2}/\sqrt{3}$ [43]. The polarized states of optical vortex beams are further analyzed by using circular analyzer consisting of one linear polarizer and one quarter waveplate. From **Figure 2** it is observed that left- and right-circularly polarized (LCP and RCP) light are projected to ± 1$^{st}$ diffraction orders respectively with efficiency of ~ 0.12% for both F2 and F3 devices. The diffraction angles for F2 and F3 metasurfaces respectively are given by $\varphi_{2, 3} \sim \pm \arcsin (\lambda/P)$. In the experiment, both linearly (H) and circularly (LCP/RCP) polarized light are used to illuminate the two metasurface fork gratings. The transmitted beam with the same SAM ($\sigma = 0, \pm 1$) is only observed at zero order, and those beams with opposite polarization states ($\sigma = \pm 1$) are observed at the first order diffractions. Thus, the metasurface shows the functionality of coupling both SAM and OAM together and projecting the coupled angular momentum of light to different spatial positions through a linear phase gradient. In this way, the metasurface fork grating can generate and control the propagation of spin polarized optical vortex with simultaneously coupled SAM-OAM modes.

We further explore the optical spin-orbital interaction when the optical vortex beam passes through the metasurface fork grating. As is shown in **Figure 3**, the first step is to generate the optical vortex beam with designed topological charges by using a binary fork mask, which is fabricated via photolithography method and wet etching process. This binary fork mask with topological charge $q = 1$ is made of 100 nm thick chromium layer on glass substrate. The fork mask has grating period of 2.4 µm with filling factor of 0.5. The He-Ne laser beam with Gaussian beam profile is focused onto the center of the fork mask by the convex lens L1 with focal length of f1 = 10 cm. The optical vortex beams with topological number $l = \pm 2$ are then generated at the second diffraction orders. The collection lens L2 is then used to transform the diffracted beam into an optical vortex with low divergence angle. Subsequently the optical vortex beam with topological number $l = -2$ is incident onto the F2 and F3 metasurface fork gratings via an objective lens with low numerical aperture: N. A. = 0.18. The transmitted light is collected by the second objective lens with N. A. = 0.4. The polarized states of transmitted light are analyzed by using linear and circular analyzers and the intensity distribution of the diffracted beams are captured by a color CCD camera. The experimental results are summarized in **Figure 4**, for H-polarized incident optical vortex $(l_{in} = -2)$, an array of OAM beams with three different topological numbers are obtained at different diffraction order for both F2 and F3 metasurface fork gratings. It is found that the topological charge of the metasurface fork does not change the OAM of incident light at zero order diffraction direction. By using metasurface fork grating F2, the incident vortex beam is converted to two new SAM-OAM coupled beams with topological number of $l_{+1} = 0$ (no singularity at the center) at +1st order and $l_{-1} = -4$ at -1st diffraction orders. It is expected that one can easily manipulate the OAM values of diffracted light by introducing various topological values into the metasurface fork grating. This idea is further verified by using F3 metasurface, which contains a topological charge $q = 3$. We obtain the left circularly polarized vortex beam with $l_{+1}= 1$ and right circularly polarized vortex beam with $l_{-1} = -5$, respectively. All the above measurements focus on the first-order diffractions on the fork gratings, which are more efficient than that from higher orders. The topological number of the generated vortex beam can be predicted by a simple formula: $l_{out} = l_{in} \pm q$, where $l_{in}$ is the topological number of input light and $q$ is the topological charge of the metasurface fork grating. In terms of polarization beam splitting effect, we can see that all the diffracted light at $\pm 1^{st}$ orders are left- and right- circularly polarized respectively no matter what the OAM states are. This phenomenon is very similar to our observations in **Figure 2**, and it means the linear phase gradient in the metasurface not only works for Gaussian beam but also for optical vortex beam as well.

### 3. Conclusion and Outlook

In conclusion, we have developed the metasurface fork gratings, made of subwavelength thick aluminum film, by combining phase singularity of the fork pattern and linear phase gradient of the metasurface together. This metasurface fork grating can act as both a polarization beam splitter and an OAM generator for Gaussian beam. Taking the advantage of strong optical spin-orbit interaction on the metasurface, one can convert the OAM states of an incident optical vortex to new modes by adding additional topological charges of the metasurface, while SAM states of light determines their propagation directions. The proposed metasurface fork grating provides a novel chip-integration platform for OAM based high capacity optical communication applications, including information coding, multiplexing, de-multiplexing of light and so on.

**Acknowledgements** S. M. and Y. C. contributed equally to this work. Y. C. thanks the support from SEG_HKUST08; S. Z was supported EPSRC. K. W. would like to thank the support from Research Grant Council of Hong Kong under Project AoE/P-02/12 and .

Figures

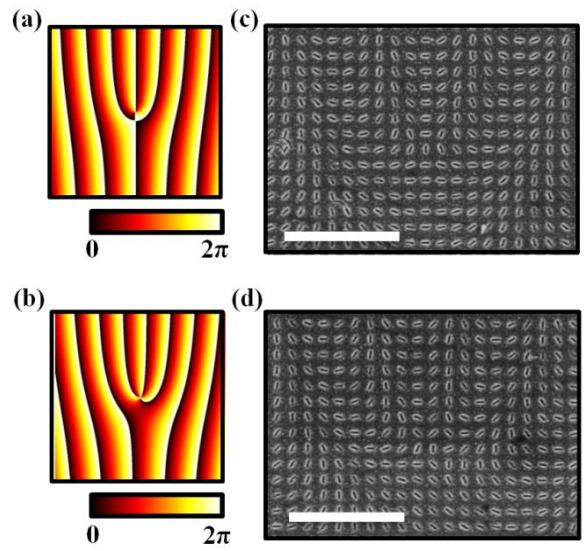

**Figure 1**. (a) and (b) Phase distribution of metasurface fork gratings with topological charge of q =2, 3; (c) and (d) Scanning electron images of plasmonic metasurfaces (F2 and F3), which wasfabricated on 80 nm thick aluminum thin film by using focus ion beam method, , consists of spatially variant nanoslits with size of : ~ 50 nm by 210 nm. Scale bar: 3 μm.

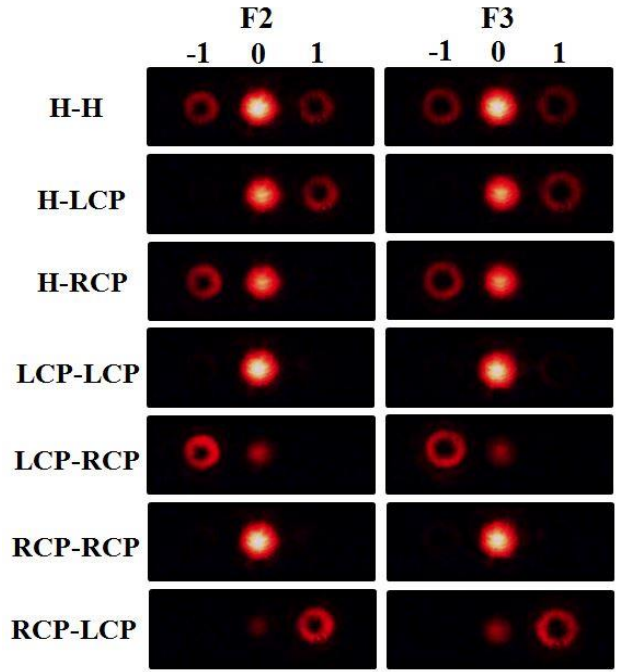

**Figure 2**. Generated optical vortex beam from metasurface fork gratings: FF2 and FF3 s under illumination laser with Gaussian beam profile. The optical vortex with SAM value: $\sigma = \pm 1$ and OAM value: $l = \pm 2, \pm 3$ are obtained at $\pm 1^{st}$ diffraction orders. H means that the incident or transmitted laser has horizontal polarization ; LCP and RCP are the polarization states of incident or transmitted beams.

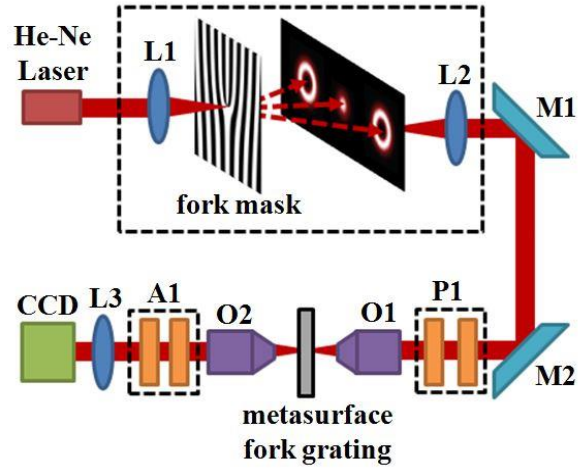

**Figure 3**. Experimental setup for studying the interaction between optical vortex and metasurface fork grating. The Gaussian beam profile of He-Ne laser is converted to optical vortex with OAM value: $l = \pm$ n (diffraction orders) and SAM: $\sigma = 0, \pm 1$ after passing through the binary fork mask made of chromium on glass substrate. The second diffraction order ($l$=-2) of transmitted light was focused onto the metasurface fork gratings through a microscope setup (O1 and O2). Both zero and $\pm 1$ diffraction orders are projected onto the CCD screen. Polarizer P1 and analyzer A1 are used to control and analyze the polarization states of the incident and transmitted light.

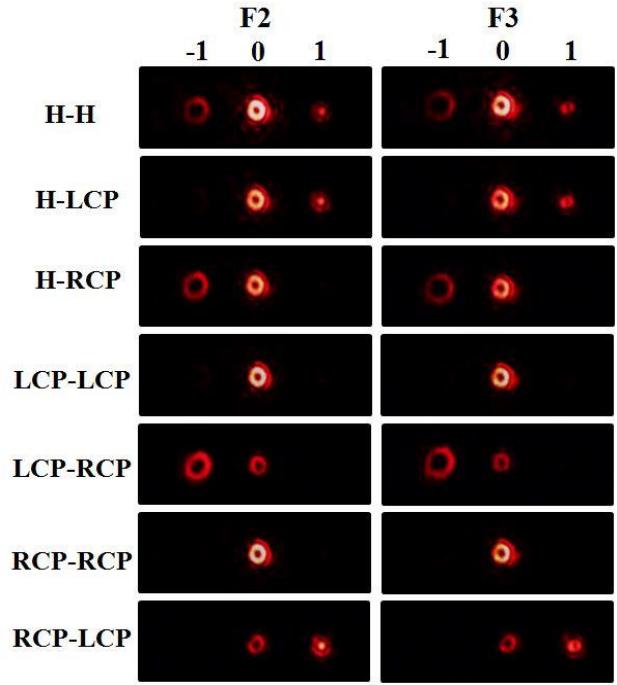

**Figure 4**. CCD imaging of interaction between optical vortex and metasurface fork gratings (F2 and F3). F2F3Incident optical vortex has orbital angular momentum (OAM) value of $l = -2$. The transmitted light from F2 metasurface fork grating generate SAM-OAM modes of light with SAM value: $\sigma = \pm 1$ and OAM value: $l = 0$, -4 at $\pm 1^{st}$ diffraction orders.in comparison, the F3 metasurface generate OAM values $l=1$, -5 are $\pm 1^{st}$ diffraction directions. H means that the incident or transmitted laser has horizontal polarization ; LCP and RCP are the polarization states of incident or transmitted beams.